\documentclass[aps,prb,twocolumn,showpacs,amsfonts,superscriptaddress]{revtex4-2}
\usepackage{epsfig}\usepackage{subfigure}
\usepackage{graphicx,graphics}
\usepackage{amsmath}
\usepackage{amssymb}
\usepackage{bm}
\usepackage{color}
\usepackage[colorlinks,linkcolor={blue},citecolor={blue},urlcolor={blue}]{hyperref}
\usepackage[utf8]{inputenc}

\begin{document}

\title{Local density of state oscillations in laterally heterostructured topological insulator-semiconductor systems}

\author{David J. Alspaugh}
\affiliation{Department of Physics and Astronomy, California State University, Northridge, California 91330, USA}
\affiliation{Sorbonne Universit{\'e}, CNRS, Laboratoire de Physique Th{\'e}orique de la Mati{\`e}re Condens{\'e}e, LPTMC, F- 75005 Paris, France}

\author{D. N. Sheng}
\affiliation{Department of Physics and Astronomy, California State University, Northridge, California 91330, USA}

\author{Mahmoud M. Asmar}
\affiliation{Department of Physics, Kennesaw State University, Marietta, Georgia 30060, USA}

\date{February 12, 2023}

\begin{abstract}

{We study local density of state (LDOS) oscillations arising from the scattering of electrons at atomic edge defects in topological insulator (TI) surfaces. To create edge scattering on the surface of a TI, we assume that half of its surface is covered with a semiconductor. In addition to modifying the TI states in the covered half, the presence of the semiconductor leads to a localized edge potential at the vacuum-semiconductor boundary. We study the induced LDOS by imposing time-reversal (TR) invariance and current conservation across the boundary. Additionally, we explore how the scattering of TI junctions with dissimilar spin textures and anisotropic Fermi velocities affect the modulations of the LDOS away from the junction edge. In all cases, for energies close to the Dirac point, we find that the decay envelope of the LDOS oscillations is insensitive to the scattering at the atomic edge defect, with a decay power given by $x^{-3/2}$. Quantitative differences in the amplitude of these oscillations depend on the details of the interface and the spin textures, while the period of the oscillations is defined by the size of the Fermi surface.}

\end{abstract}

\pacs{}

\maketitle

\section{Introduction}
Topological surface states have attracted much attention due to their potential applications in quantum computing and spintronics~\cite{Hasan2010,Qi2011}. These novel electronic states have been observed on the surfaces of three-dimensional topological insulators (TIs) such as Bi$_{2}$Se$_{3}$~\cite{Xia2009,Hsieh2009}, Bi$_{1-x}$Sb$_{x}$~\cite{Hsieh2008,Hsieh2009a,Roushan2009}, Sb$_{2}$Te$_{3}$~\cite{Zhang2009a}, and Bi$_{2}$Te$_{3}$~\cite{Chen2009,Zhang2009a} through angle-resolved photoemission spectroscopy. The band structure of these topologically protected bound surface states was originally determined by employing $k\cdot p$ perturbation theory and at low energies they possess a helical Dirac-type dispersion. In contrast, the topologically protected interface states found in junctions of TIs with topologically trivial materials such as semiconductors (SEs) can possess more exotic nonhelical spin textures with anisotropic Fermi velocities and spins that point out of the plane of the interface~\cite{Alspaugh2018,Asmar2017,Thareja2020,Brems2018}.

The helical nature of the bound surface states implies that the backscattering of these states by time-reversal (TR) invariant impurities is greatly suppressed~\cite{Roushan2009}. The consequences of this suppression can be seen in scanning tunneling microscopy experiments, where the quasiparticle scattering arising from the presence of a surface step defect can lead to the observation of oscillations in the local density of states (LDOS)~\cite{Zhang2009,Gomes2009,Alpichshev2010,Alpichshev2011,
Wang2011}. While the LDOS oscillations about an edge defect decay as $x^{-1/2}$ in a conventional two-dimensional electron gas (2DEG), with $x$ being the distance from the edge, the LDOS oscillations on TIs with energy close to the Dirac point are found to decay as $x^{-3/2}$~\cite{Biswas2011,Wang2011,Liu2012}. For Bi$_2$Te$_3$, energies closer to the band gap edge lead to hexagonal warping effects, resulting in longer LDOS oscillation decay envelopes such as $x^{-1}$~\cite{Zhou2009,Wang2011,Rakyta2012,Zhang2012,An2012}. However, most theoretical models have so far focused on systems where both sides of the edge defect either have the same helical spin texture, or have mismatched Fermi velocities~\cite{McKellar1987,Takahashi2011,Sen2012,Shao2014}.

\begin{figure}
\includegraphics[width=\columnwidth]{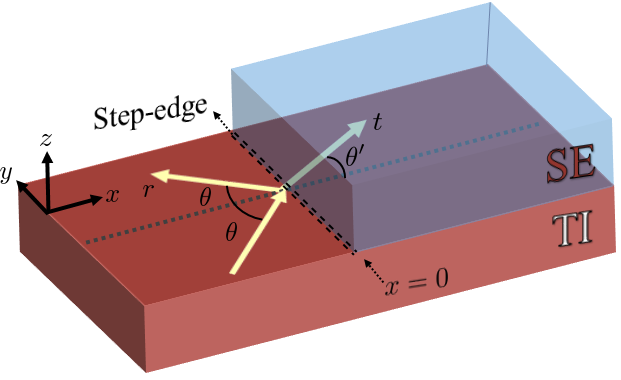}
\caption{Schematic of an atomic step-edge created by placing a semiconductor (SE) over half of a topological insulator (TI) surface. Here we set $x = 0$ as the location of the ``step-edge" interface between the TI-vacuum and TI-SE regions. Arrows indicate the possible outcomes of quasiparticle scattering at the interface. An incident electron from the TI-vacuum surface approaches the interface with an angle $\theta$. The electron can either be reflected back towards the TI-vacuum region as shown by $r$, or can be transmitted through the interface to the TI-SE region with angle $\theta'$ as given by $t$.}
\label{schem}
\end{figure}

In this work we study the LDOS oscillations about step-edges that emerge from placing a SE over half of a TI surface, as shown schematically in Fig.~\ref{schem}. This hybrid system is characterized by two main features: Firstly, in the covered half, TR preserving surface boundary effects such as lattice strain, dangling bonds at the interface, and charge accumulation can lead to the formation of elliptical energy contours and nonhelical spin textures in the interface states, in contrast to the helical surface states of the uncovered half~\cite{Zhang2012a,Brems2018}. Secondly, at the one-dimensional boundary between the TI-vacuum and TI-SE regions, ``step-edge" scattering potentials that are related to the presence of Dirac delta-like interface potentials emerge. Both of these effects result in quasiparticle scattering at the step-edge giving rise to LDOS oscillations that decay away from the boundary. We find that the decay envelope of these LDOS oscillations is robust against the details of the step-edge disorder, with oscillations generally decaying away from the edge as $x^{-3/2}$, while the amplitude and period of these oscillations depend on the details of the junction.

The paper is then organized as follows: In Section~\ref{sec2} we introduce the Hamiltonian of the lateral TI heterojunction shown in Fig.~\ref{schem} and join the wave functions of the TI-vacuum and TI-SE regions. Using these boundary conditions, which encode both the nonhelicity of the TI-SE region and the localized scattering potential of the step-edge, we set up the quasiparticle scattering problem of incoming electrons towards the step-edge and calculate the reflection and transmission coefficients along with the LDOS of the TI-vacuum region. In Section~\ref{sec3} we focus on the special case where both sides of the step-edge admit equal helical surface states and both analytically and numerically analyze the LDOS. In Section~\ref{sec4} we analyze how the LDOS oscillations are affected by altering the relative sizes of the Fermi surfaces on either side of the junction. In Section~\ref{sec5} we introduce examples of nonhelical spin textures and determine possible experimental signatures which distinguish them from those of the helical examples considered previously. In particular, we analyze systems that exhibit arbitrarily oriented elliptical Fermi surfaces with spin textures that point out of the plane of the TI-SE planar interface. Finally in Section~\ref{sec6} we discuss the experimental consequences of step-edge disorder in TI hybrid systems.

\section{Boundary Conditions}
\label{sec2}
To study the LDOS of the TI-SE lateral heterostructure shown in Fig.~\ref{schem} we define our coordinate system such that the the $x-y$ plane constitutes the surface of the TI, where the $x<0$ half space contains the TI-vacuum surface and the $x>0$ half space contains the TI-SE interface, with the boundary between these two regions being along $x = 0$. While the TI-vacuum surface hosts a helical TI surface state with spins confined to the surface, the material junction of the TI-SE planar interface can have interface states with nonhelical spin textures, giving rise to elliptical constant energy contours and spins with components that point out of the plane of the interface (i.e. along the $z$-direction). To study this heterojunction we consider a helical surface Hamiltonian on the left and the most general effective linear Hamiltonian that is TR invariant on the right,
\begin{equation}
\begin{split}
H &= H_{\rm L}\Theta(-x) + H_{\rm R}\Theta(x),
\\ H_{\rm L} &= \hbar v_{\rm F}(\bm{\sigma}\times -i\bm{\nabla})_{z} - \mu_{\rm L},
\\ H_{\rm R} &= \bm{c}\cdot\bm{\sigma} - \mu_{\rm R}.
\end{split}
\label{Ham}
\end{equation}
Here $\Theta(x)$ is the Heaviside step function, $\bm{\sigma} = (\sigma_{x},\sigma_{y},\sigma_{z})^{T}$ is a vector of Pauli matrices in spin space, $v_{\rm F}$ is the Fermi velocity of the TI-vacuum surface, $\mu_{\rm L}$ and $\mu_{\rm R}$ are the left and right region's chemical potentials respectively, and $\bm{c}$ is a three component vector defined by $c_{i} = -i\sum_{j=x,y}c_{ij}\partial_{j}$. The Hamiltonian $H_{\rm R}$ preserves TR symmetry so long as the $c_{ij}$ coefficients are all real. From this description the helical TI surface state (the case in which there is no SE) may be modeled by setting $\bm{c}_{\rm D} = -i\hbar v_{\rm F}(\partial_{y},-\partial_{x},0)^{T}$. Examples of nonhelical spin textures may be found in Refs.~\cite{Zhang2012a,Asmar2017,Alspaugh2018,Brems2018,Thareja2020,Alspaugh2022},
and are thus encoded in the choice of the $c_{ij}$ coefficients.

Because we are primarily interested in low-energy conductance signatures, in Eq.~\eqref{Ham} we neglect the typical $k^{3}$ momentum terms that give rise to the hexagonal warping of the Fermi surface~\cite{Fu2009}. Without loss of generality, we may write the eigenstates of the Hamiltonian in Eq.~\eqref{Ham} as $\psi(x) = \Theta(-x)\psi_{\rm L}(x) + \Theta(x)\psi_{\rm R}$. Consequently, the effective long wavelength description of Eq.~\eqref{Ham} cannot account for the rapid variations of the wave function in the vicinity of the SE edge at $x = 0$. The proper TR preserving boundary condition for the wave function at the $x = 0$ interface is given by $\psi_{\rm L}(0) = \mathcal{M}(\beta)\psi_{\rm R}(0)$, where~\cite{Alspaugh2018,Alspaugh2022,Asmar2017}
\begin{equation}
\mathcal{M}(\beta) = \sqrt{\dfrac{v}{v_{\rm F}}}\bigg[e^{i\sigma_{y}\beta} + \dfrac{i}{2\hbar v}(c_{xx}\sigma_{z} - c_{zx}\sigma_{x})e^{-i\sigma_{y}\beta}\bigg].
\label{matrixm}
\end{equation}
Here we have that $v = (\sqrt{\sum_{i=x,y,z}c_{ix}^{2}} - c_{yx})/2\hbar$ and $\beta$ is an arbitrary parameter such that $\beta \in [0,2\pi)$. For additional details on the derivation of this boundary condition, see Appendix~\ref{appendixparticlecurrent}. For illustrative purposes, we may first focus on the case that $H_{\rm L} = H_{\rm R}$; this is equivalent to writing $\bm{c} = \bm{c}_{\rm D}$ and $\mu_{\rm L} = \mu_{\rm R}$, and the boundary value matrix takes the form $\mathcal{M}(\beta) = e^{i\sigma_{y}\beta}$. For this case in particular, it was shown that $\beta$ is proportional to the strength of a Dirac delta edge potential localized at $x = 0$~\cite{Alspaugh2022,McKellar1987}. From the form of $\mathcal{M}(\beta) = e^{i\sigma_{y}\beta}$ we see that this results in a rotation of the spin's expectation value about the $y$-axis, conserving the $x$-component of the electric current across the junction. In the case that $H_{\rm R} \neq H_{\rm L}$, Eq.~\eqref{matrixm} describes the proper transformation of the wave function's spin expectation value, and the value of the parameter $\beta$ is not simply related to the strength of a localized edge potential as it can break inversion and chiral symmetries. Hence, the free parameter of the general boundary condition in Eq.~\eqref{matrixm} captures the effects of all TR symmetry allowed disorder at the step-edge.

\subsection{Local density of states of the lateral heterojunction}

We may then we calculate the LDOS oscillations that emerge due to the quasiparticle interference patterns about the junction. To do this, we must first solve the quasiparticle scattering problem about the $x = 0$ interface. We therefore consider an incoming electron from the left half space of Fig.~\ref{schem} with momentum $\bm{k}_{1} = (k_{x},k_{y})^{T}$ and in-plane momentum angle $\theta = \tan^{-1}k_{y}/k_{x}$, a reflected electron with momentum $\bm{k}_{2} = (-k_{x},k_{y})^{T}$, and a transmitted electron in the right half space with momentum $\bm{k}_{1}' = (k_{x}',k_{y})^{T}$. The outgoing angle $\theta'$ is solved for in terms of the incoming angle due to conservation of energy and $k_{y}$ momentum. The wave function of $H_{\rm L}$ is given by
\begin{equation}
\psi_{\rm L}(\bm{r}) = \dfrac{e^{i\bm{k}_{1}\cdot\bm{r}}}{\sqrt{2}}\begin{pmatrix}
1 \\ -ie^{i\theta}
\end{pmatrix} + r \dfrac{e^{i\bm{k}_{2}\cdot\bm{r}}}{\sqrt{2}}\begin{pmatrix}
1 \\ ie^{-i\theta}
\end{pmatrix}.
\end{equation}
Similarly, the wave function of $H_{\rm R}$ is given by
\begin{equation}
\psi_{\rm R}(\bm{r}) = te^{i\bm{k}_{1}'\cdot\bm{r}}\begin{pmatrix}
\cos(\vartheta_{\bm{c}(\bm{k}_{1}')}/2)
\\ e^{i\varphi_{\bm{c}(\bm{k}_{1}')}}\sin(\vartheta_{\bm{c}(\bm{k}_{1}')}/2)
\end{pmatrix}.
\label{transmittedwavefunction}
\end{equation}
Here $\vartheta_{\bm{c}(\bm{k}_{1}')}$,$\varphi_{\bm{c}(\bm{k}_{1}')}$ are the polar and azimuthal angles of the vector $\bm{c}(\bm{k}_{1}')$ respectively, where $\bm{c}(\bm{k})$ is defined by $c_{i}(\bm{k}) = \sum_{j=x,y}c_{ij}k_{j}$, and $r$ and $t$ are the coefficients for the reflected and transmitted parts of the wave function respectively. Matching the solutions at $x = 0$ [Eq.~\eqref{matrixm}] we may solve for both the reflection and transmission coefficients, for details see Appendix~\ref{Appendixscatteringdetails}. Using these wave functions we can then calculate the LDOS from the spectral function of the exposed TI surface. The LDOS then takes the form
\begin{equation}
\begin{split}
\rho(\bm{r},\omega) &= \dfrac{1}{(2\pi)^{2}}\int_{0}^{\infty} dk_{x}\int_{-\infty}^{\infty}dk_{y}|\psi_{\rm L}(\bm{r})|^{2}\delta(\omega - \xi_{\bm{k}})
\\ &= \dfrac{\rho_{0}}{2\pi}\int_{-\pi/2}^{\pi/2}d\theta D_{k_{\omega}\theta x}.
\end{split}
\label{rho}
\end{equation}
Here $\xi_{\bm{k}} = \hbar v_{\rm F}\sqrt{k_{x}^{2} + k_{y}^{2}} - \mu_{\rm L}$ is the energy of the $x<0$ electrons, $k_{\omega} = \tfrac{\omega + \mu_{\rm L}}{\hbar v_{\rm F}}$ is the magnitude of the momentum at energy $\omega$, $\rho_{0} = \tfrac{k_{\omega}}{2\pi\hbar v_{\rm F}}$ is the constant LDOS of a single TI surface slab with no semiconducting material on the $x>0$ side, and the integrand is given by
\begin{equation}
\begin{split}
D_{k \theta x} &= 1 + |r(\theta)|^{2} + \text{Re}[e^{-2ik\cos\theta x}r(\theta)]
\\ &\hspace{2cm} - \text{Re}[e^{-2ik\cos\theta x}e^{-2i\theta}r(\theta)].
\end{split}
\label{integrand}
\end{equation}
In the following sections we consider some specific examples of $H_{\rm R}$ which will allow us to obtain both analytic and numerical results for $r(\theta)$, $D_{k_{\omega}\theta x}$, and $\rho(\bm{r},\omega)$.

\section{step-edge scattering: equal Fermi surfaces}
\label{sec3}

\begin{figure}
\includegraphics[scale=0.6]{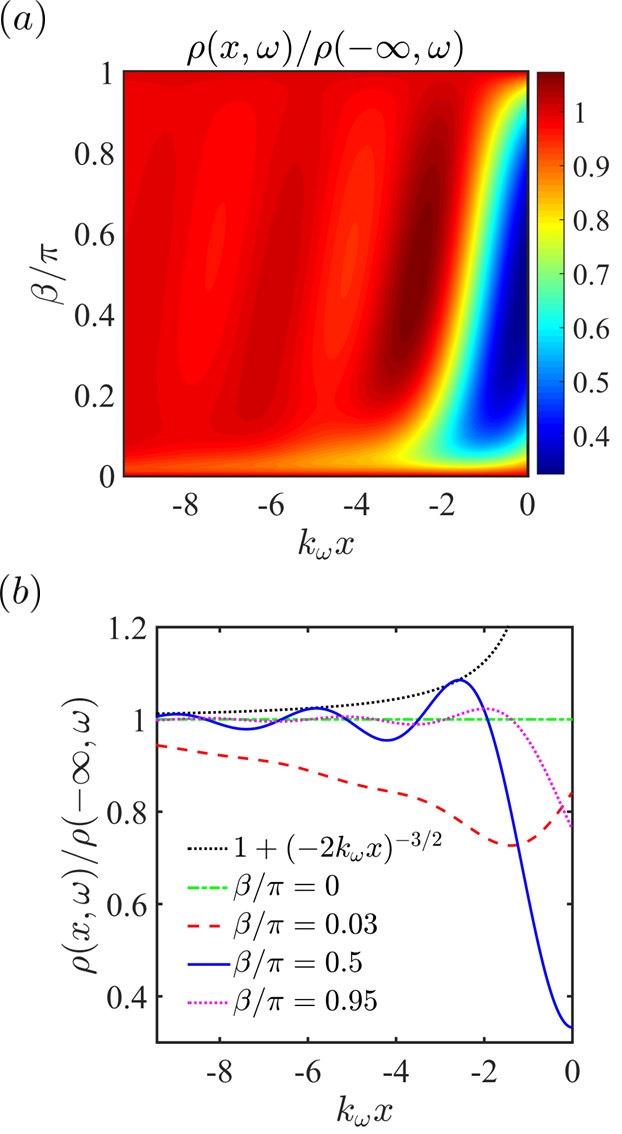}
\caption{LDOS oscillations of the TI-vacuum region shown in Fig.~\ref{schem}. Here the TI-SE region is assumed to have the same surface Hamiltonian of the TI-vacuum region. (a) LDOS oscillations given as a function of position and $\beta$, where in this case $\beta$ has been shown to be proportional to the strength of an edge potential localized at the $x = 0$ interface. The LDOS is normalized by its value as $x\to-\infty$, which need not be the same for each value of $\beta$. (b) Examples of the LDOS profile are plotted for chosen values of $\beta$, along with the $x^{-3/2}$ decay envelope.}
\label{fig2}
\end{figure}

We first analyze the LDOS oscillations when both regions of our junction admit equal helical TI surface states. As previously described, this case can be described within our model by setting $\bm{c} = \bm{c}_{\rm D}$ and $\mu_{\rm L} = \mu_{\rm R}$, and the boundary value matrix of Eq.~\eqref{matrixm} has the form $\mathcal{M} = e^{i\sigma_{y}\beta}$, where the free parameter $\beta$ is proportional to the strength of a localized edge potential. By solving the scattering problem shown schematically in Fig.~\ref{schem} we find the reflection coefficient of the reflected electrons to have the form
\begin{equation}
r(\theta,\beta) = e^{i\theta}\dfrac{\sin\theta}{-i + \cos\theta\cot\beta}.
\label{equalcircler}
\end{equation}
This expression may also be found by setting $\theta' = \theta$ in Eq.~\eqref{rviv2} of Appendix~\ref{appendixreflection}. In Fig.~\ref{fig2} we numerically integrate Eq.~\eqref{rho} and plot the LDOS as a function of position and edge potential strength $\beta$. Unique LDOS profiles are observed for $\beta\in[0,\pi)$, beyond which they become periodic. As expected, a constant LDOS profile is observed in the absence of any edge potential due to the lack of scattering. This is evident as when $\beta$ vanishes, we have that $r(\theta) = 0$ and $D_{k_{\omega}\theta x} = 1$, giving rise to $\rho(x,\omega) = \rho_{0}/2$. As the edge potential strength increases, oscillations develop and increase due to electrons scattering at the edge and a minimum in the global local density occurs at a finite distance away from the $x = 0$ interface. The oscillations are largest when $\beta = \pi/2$, which corresponds to the largest strength of the edge potential, and in this case the minimum of the LDOS profile approaches $x = 0$. As the potential strength increases in the region $\beta\in[\pi/2,\pi)$ the oscillations diminish, although the minimum of the global local density remains fixed at $x = 0$.

In the special case that $\beta = \pi/2$ the LDOS can be expressed analytically. From Eq.~\eqref{equalcircler} the reflection coefficient is $r(\theta) = ie^{i\theta}\sin\theta$, and the integrand of the LDOS is $D_{k_{\omega}\theta x} = 1 + \sin^{2}[1 - 2\cos(2k_{\omega}x\cos\theta)]$. Using the Jacobi-Anger expansion we can then calculate the LDOS
\begin{equation}
\rho(x,\omega) = \rho_{0}\bigg[\dfrac{3}{4} - \dfrac{J_{1}(2k_{\omega}x)}{2k_{\omega}x}\bigg].
\label{equalcirclerho}
\end{equation}
Here $J_{n}(z)$ in the $n$th Bessel function of the first kind. Because the Bessel functions asymptotically decay as $z^{-1/2}$ we can see that the second term in $\rho(x,\omega)$ decays as $z^{-3/2}$. Then for $x\to-\infty$ we find $\rho(-\infty,\omega) = 3\rho_{0}/4$. Consequently, $\rho(x,\omega)/\rho(-\infty,\omega)$ asymptotically decays as $1 + (-2k_{\omega}x)^{-3/2}$ as shown in Fig.~\ref{fig2}(b) (recall that $x<0$). The oscillations of the quasiparticle interference pattern have a wave length determined by $\pi/k_{\omega}$. From the interpretation that $\beta$ controls the strength of a localized edge potential at the $x=0$ interface (Fig.~\ref{schem}), we may observe that increasing the strength of this edge potential increases the amplitude of the LDOS oscillations until they reach their maximum at $\beta = \pi/2$. After this, the oscillations must decrease with increasing potential strength in order to maintain the periodicity of $\beta$ in the interval $\beta\in[0,\pi)$. The suppression of backscattering in TI junctions due to the helical nature of their charge carriers, and the fact that the boundary conditions encoded by the matrix $\mathcal{M}(\beta)$ relate incoming and outgoing quasiparticles through rotations of their spin expectation value at the $x = 0$ step edge, leads to LDOS oscillations that decay as $x^{-3/2}$, which is qualitatively shorter than those mediated by nonhelical Schr\"{o}dinger electrons in 2DEGs~\cite{Biswas2011,Wang2011,Liu2012}.

\section{step-edge scattering: unequal Fermi surfaces}
\label{sec4}

\begin{figure}
\includegraphics[scale = 0.375]{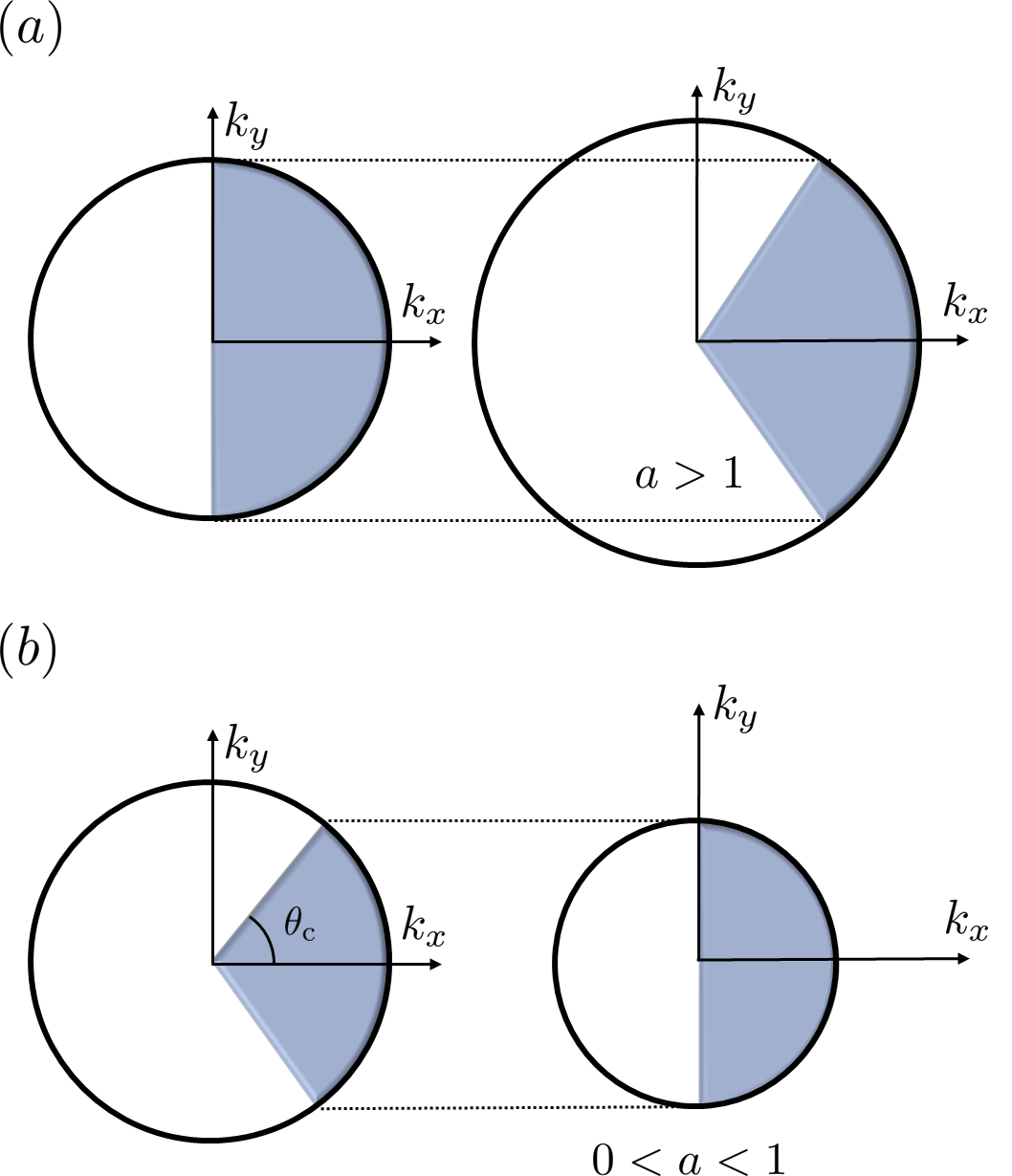}
\caption{Schematics of the Fermi surfaces of the TI-vacuum surface (left) and the TI-SE interface (right). The relative sizes of the two Fermi surfaces are controlled through the parameter $a = \tfrac{\mu_{\rm R}}{v_{\rm R}}\tfrac{v_{\rm L}}{\mu_{\rm L}}$. (a) When $a > 1$, the Fermi surfaces of the TI-SE region is larger than the TI-vacuum region. (b) When $0 < a < 1$, the Fermi surface of the TI-SE region is smaller than that of the TI-vacuum region. In this case, incident electrons from the TI-vacuum region can only be transmitted into the TI-SE region if they have an incoming angle less than the critical angle $\theta_{\rm c} = \arcsin a$. }
\label{fermischemfig}
\end{figure}

\begin{figure*}
\includegraphics[width=\textwidth]{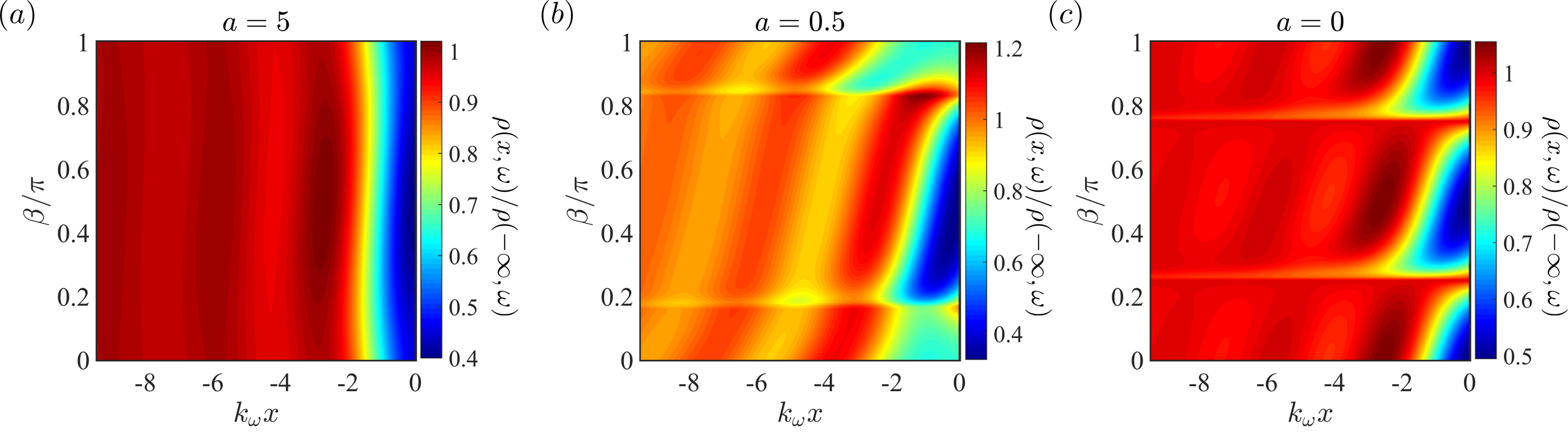}
\caption{LDOS as a function of $\beta$ in the case of unequal Fermi surfaces across the junction. Here the parameter $a = \tfrac{\mu_{\rm R}}{v_{\rm R}}\tfrac{v_{\rm L}}{\mu_{\rm L}}$ represents the ratio of the relative sizes of the Fermi surfaces. When $a>1$ ($a<1$) the Fermi surface of the TI-SE interface is larger (smaller) than that of the TI-vacuum surface.}
\label{figatable}
\end{figure*}

\begin{figure}
\includegraphics[scale=0.35]{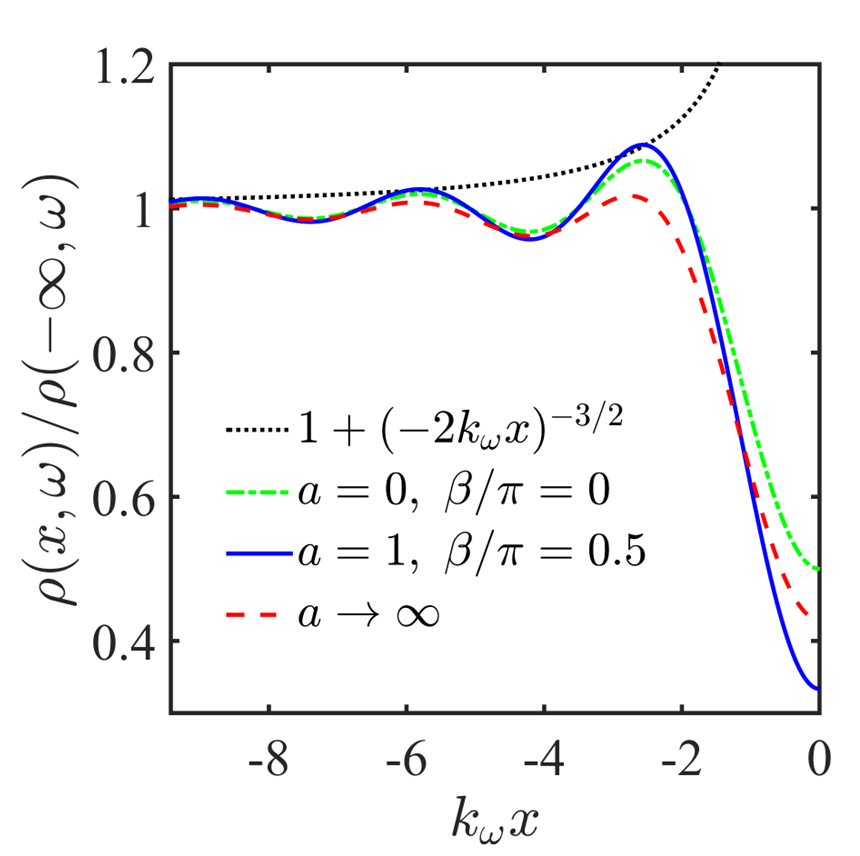}
\caption{Comparison of the analytical solutions for the maximum LDOS oscillations in the different limits of $a$.}
\label{figasmall}
\end{figure}

In this section we once again consider the case that both regions of the junction admit helical TI surface states. However, this time we allow the TI-SE interface to alter the size of the Fermi surface. This can be accomplished either by changing the chemical potential or by changing the Fermi velocity on the $x>0$ side of the junction. To study the scattering of TI surface states with unequal Fermi surfaces, we may rewrite the junction described in Eq.~\eqref{Ham} and consider the Hamiltonian
\begin{equation}
\begin{split}
H_{\rm L} &= \hbar v_{\rm L} (\sigma_{x}k_{y} + i\sigma_{y}\partial_{x}) - \mu_{\rm L},
\\ H_{\rm R} &= \hbar v_{\rm R} (\sigma_{x}k_{y} + i \sigma_{y}\partial_{x}) - \mu_{\rm R}.
\end{split}
\label{BigcircleSmallcircleHam}
\end{equation}
Here $v_{\rm L}$, $\mu_{\rm L}$ and $v_{\rm R}$, $\mu_{\rm R}$ are the Fermi velocities and the chemical potentials of the left and right sides of the junction respectively. In this setup the boundary value matrix of Eq.~\eqref{matrixm} now has the form $\mathcal{M}(\beta) = \sqrt{v_{\rm R}/v_{\rm L}}e^{i\sigma_{y}\beta}$. Before solving the scattering problem to find the reflection coefficient and the resulting LDOS oscillations, we first must calculate the outgoing angle $\theta'$ in terms of the incoming angle and system parameters. Due to translational invariance along the $y$-direction, the conservation of the $k_{y}$ momentum yields the condition
\begin{equation}
\dfrac{\mu_{\rm L}}{v_{\rm L}}\sin\theta = \dfrac{\mu_{\rm R}}{v_{\rm R}}\sin\theta',
\label{snellslaw}
\end{equation}
which is analogous to Snell's law in optics. Defining the parameter $a = \tfrac{\mu_{\rm R}}{v_{\rm R}}\tfrac{v_{\rm L}}{\mu_{\rm L}}$, the outgoing angle is then $\theta' = \arcsin(a^{-1}\sin\theta)$.

The value of $a$ determines the relative sizes of the Fermi surfaces. As shown schematically in Fig.~\ref{fermischemfig}, when $a > 1$ the Fermi surface of the TI-SE interface is larger than that of the TI-vacuum surface. Conversely, when $0 < a < 1$ the Fermi surface of the TI-SE interface is smaller than that of the incident side. In this case, electrons can only propagate into the TI-SE region as long as the incoming angle is less than the critical angle $\theta_{\rm c} = \arcsin a$. For incoming angles greater than $\theta_{\rm c}$, total internal reflection occurs and there are only decaying modes within the TI-SE region.

At this level we can numerically obtain the LDOS profile using Eq.~\eqref{rho}. We note, however, that the quasiparticle scattering problem is slightly modified for angles greater than $\theta_{\rm c}$, as the wave functions $\psi_{\rm R}(x)$ described by Eq.~\eqref{transmittedwavefunction} must instead be replaced by evanescent modes. In Fig.~\ref{figatable} we plot the LDOS as a function of the TR invariant edge perturbations for three sets of Fermi surfaces with different relative sizes, such that $a = 5$, $0.5$, and $0$. In Fig.~\ref{figatable}(a), with $a = 5$, the Fermi surface of the TI-SE interface is larger than that of the TI-vacuum surfaces. As before, the LDOS oscillations as a function of position are still present, and are periodic with edge potential strength in the interval $\beta\in[0,\pi)$. However, there is less variability in the LDOS profiles, and oscillations are now always present. In the extreme scenario when $a\to\infty$, the variability is gone and we obtain a single LDOS profile independent of the strength of the edge potential. In Fig.~\ref{figatable}(b), when $a = 0.5$, the Fermi surface of the TI-SE interface is now smaller than the TI-vacuum surface's Fermi surface. In this case as well, LDOS oscillations are now always present, and the system is periodic in the interval of length $\pi$. Finally, in Fig.~\ref{figatable}(c) when $a = 0$, i.e., when the right-hand Fermi surface vanishes, the LDOS spectra suddenly and uniquely become periodic in the smaller interval of length $\pi/2$. For $\beta = \pm\pi/4$ the LDOS is constant and the oscillations vanish, while the $\beta\in(-\pi/4,\pi/4)$ region strongly resembles the LDOS spectra obtained in Section~\ref{sec3} for equal Fermi surfaces. For the cases when $a < 1$ we observe that the presence of evanescent modes in the TI-SE region does not affect the $x^{-3/2}$ decay envelope of the LDOS oscillations.

Additional physical insights can be gained when considering the limiting values of the parameter $a$ which can be analytically solved for. Let us first study the $a \to \infty$ case (i.e., when $\mu_{\rm R} \gg \mu_{\rm L}$ or $v_{\rm L} \gg v_{\rm R}$), in which the TI-SE Fermi surface is much larger than the incident side. From Snell's law in Eq.~\eqref{snellslaw}, we see that the outgoing angle is always $\theta' = 0$ and the reflection coefficient becomes
\begin{equation}
r(\theta) = ie^{i\theta}\tan(\theta/2).
\label{abigref}
\end{equation}
This expression may also be found by setting $\theta' = 0$ in Eq.~\eqref{rviv2} of Appendix~\ref{appendixreflection}. Note that in this case the reflection coefficient, and thus the LDOS, is independent of the free parameter $\beta$ controlling the edge potential strength in the boundary value matrix. To understand this, recall from Section~\ref{sec2} that in the presence of a localized edge potential the spin of the transmitted electrons is rotated about the $y$-axis by an angle of $\beta$. Generally, every incoming state with angle $\theta$ corresponds to a transmitted state with angle $\theta'$. However, in the $a \to \infty$ limit, all outgoing states are fixed to have an outgoing angle of $\theta' = 0$. From the Hamiltonian in Eq.~\eqref{BigcircleSmallcircleHam}, we can see that the spin of these outgoing states is similarly fixed to point directly along the $y$-axis. Therefore these states are unaffected by the rotations arising from the localized edge potential, as demonstrated in the reflection coefficient in Eq.~\eqref{abigref}.

The integrand of the LDOS is $D_{k_{\omega}\theta x} = 1 + \tan^{2}(\theta/2) - 4\cos(2k_{\omega}x\cos\theta)\sin^{2}(\theta/2)$. Using the Jacobi-Anger expansion the LDOS may be integrated to give
\begin{equation}
\begin{split}
\rho(x,\omega) &= \dfrac{2}{\pi}\rho_{0} - \bigg(1 - \dfrac{2}{\pi}\bigg)\rho_{0}J_{0}(2k_{\omega}x)
\\ &\hspace{2cm} - \dfrac{4}{\pi}\rho_{0}\sum_{n = 1}^{\infty}\dfrac{J_{2n}(2k_{\omega}x)}{4n^{2} - 1}.
\end{split}
\label{bigcirclerho}
\end{equation}
For $x \to -\infty$ we find $\rho(-\infty,\omega) = 2\rho_{0}/\pi$. Moreover, asymptotically expanding $\rho(x,\omega)$ up to order $x^{-3/2}$ we notice that the $x^{-1/2}$ terms are equal and opposite in sign, rendering $x^{-3/2}$ as the leading decay power of $\rho(x,\omega)$.

We also find an analytic solution for the opposite limit where $a = 0$. In this case the Fermi surface of the TI-SE region is vanishingly small, and the critical angle is $\theta_{\rm c} = 0$. Because of this, electrons with any arbitrary angle are reflected in what mimics total internal reflection in optics. Here the scattering problem must be solved again with the assumption that there are only evanescent states in the $x>0$ side, modifying the wave function of Eq.~\eqref{transmittedwavefunction}. The reflection coefficient is given by
\begin{equation}
r(\theta,\beta) = -\dfrac{\cos\beta + ie^{i\theta}\sin\beta}{\cos\beta - ie^{-i\theta}\sin\beta},
\end{equation}
and naturally $|r(\theta,\beta)|^{2} = 1$, consistent with total internal reflection. In this limit the LDOS can be analytically derived for multiple values of the edge potential strength $\beta$. When $\beta = \pi/4$, it can be demonstrated that the LDOS has a constant profile, $\rho(x,\omega) = \rho_{0}$. The largest LDOS oscillations occur when $\beta = 0$ and $\beta = \pi/2$, which both have the same analytic solution for the LDOS given by
\begin{equation}
\rho(x,\omega) = \rho_{0}\bigg[1 - \dfrac{J_{1}(2k_{\omega}x)}{2k_{\omega}x}\bigg].
\label{smallcirclerho}
\end{equation}
For $x\to -\infty$ we find $\rho(-\infty,\omega) = \rho_{0}$. Once again, the decay power of this LDOS profile is given by $x^{-3/2}$. The LDOS oscillation patterns of both the $a \to \infty$ and $a = 0$ limits, as given by Eq.~\eqref{bigcirclerho} and Eq.~\eqref{smallcirclerho}, are plotted in Fig.~\ref{figasmall} along with the results of the equal Fermi surface case of Section~\ref{sec3}. While both the $a = 1$ and $a = 0$ cases can give rise to constant LDOS profiles, the largest oscillations are only possible in the system with equal Fermi surfaces. In contrast, the LDOS oscillations of the $a \to \infty$ case are robust and insensitive to the Dirac delta-like scattering of the $x = 0$ interface. For all sizes of Fermi surfaces considered (that is, for all values of $a$), the decay envelope is always found to be given by $x^{-3/2}$.

\section{Consequences of nonhelical spin textures}
\label{sec5}
\begin{figure}
\includegraphics[scale=0.6]{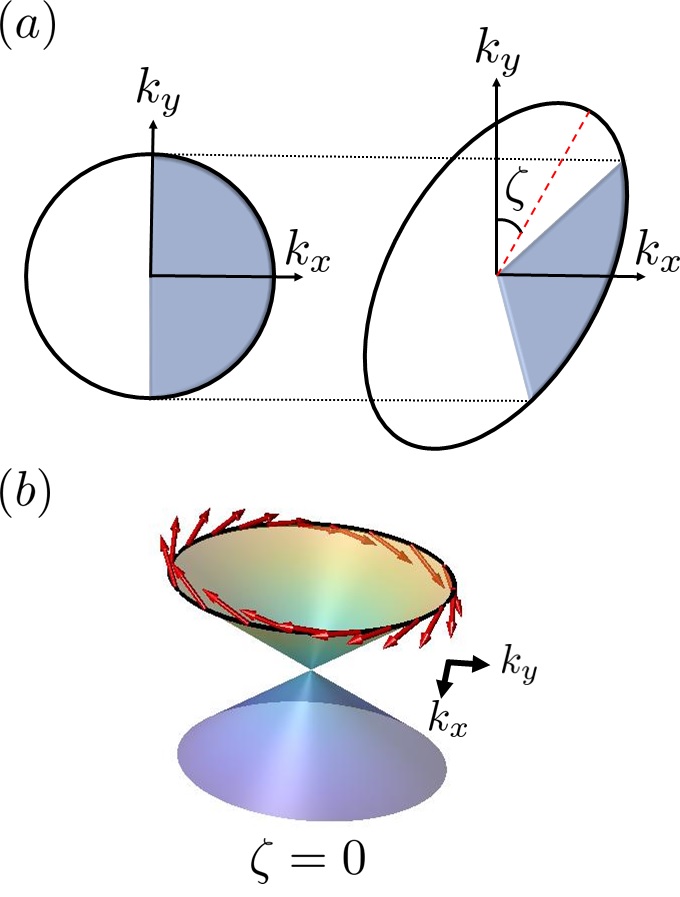}
\caption{Nonhelical dispersion and spin texture of the TI-SE interface as given by Eq.~\eqref{cperp}. (a) Schematics of the Fermi surfaces of the TI-vacuum surface (left) and the nonhelical TI-SE interface (right). The angle $\zeta$ controls the overall tilt of the elliptical Fermi surface, and is measured from the $y$-axis. (b) Spin texture of the nonhelical interface state as $\zeta = 0$. The semi-major axis of the ellitpical Fermi surface is parallel to the $x = 0$ boundary, and when $k_{y}\neq 0$, there is an out-of-plane spin component.}
\label{figelipangle}
\end{figure}

Lastly, we study how the LDOS oscillations are affected by the presence of nonhelical spin textures within the TI-SE interface. Nonhelical interface states at low energies result from inversion symmetry breaking interface potentials within TIs and topologically trivial materials, and can introduce elliptical constant energy contours in momentum space as well as out-of-plane spin textures~\cite{Alspaugh2018,Asmar2017}. To account for these types of spin textures, we may rewrite $H_{\rm R}$ in Eq.~\eqref{Ham} such that
\begin{equation}
\begin{split}
H_{\rm R} &= v_{\rm F}\bigg[({\bm \sigma}\times -i\bm{\nabla})_{z}+\dfrac{1}{3}({\bm \sigma}\cdot \bar{\bm{e}})(-i\bm{\nabla}\cdot{\bm e})
\\ &\hspace{4cm} +\dfrac{2}{3}\sigma_{z}i\bm{\nabla}\cdot{\bm e}\bigg] - \mu.
\end{split}
\label{cperp}
\end{equation}
Here, the unit vectors $\bm{e} = [\sin(\zeta), \cos(\zeta), 0]^{T}$ and $\bar{\bm{e}} = [-\cos(\zeta),\sin(\zeta),0]^{T}$ within the $x-y$ plane point along the semi-major and semi-minor axes of the elliptical Fermi surface respectively, with the angle $\zeta$ being defined from the $y$-axis, and we have set $\mu_{\rm L} = \mu_{\rm R} = \mu$. An illustration of how the Fermi surface of $H_{\rm R}$ compares to that of $H_{\rm L}$, along with the resulting spin texture of $H_{\rm R}$, are shown in Fig.~\ref{figelipangle}. In the case that $\zeta = 0$ [Fig.~\ref{figelipangle}(b)], electrons traveling along the $y$-direction have spin expectation values that contain an out-of-plane component, in contrast to the helical case studied previously. We also note that even though the nonhelical interface state defined by Eq.~\eqref{cperp} is distinct from the helical surface state, the system under consideration is still TR invariant.

\begin{figure*}
\includegraphics[width=\textwidth]{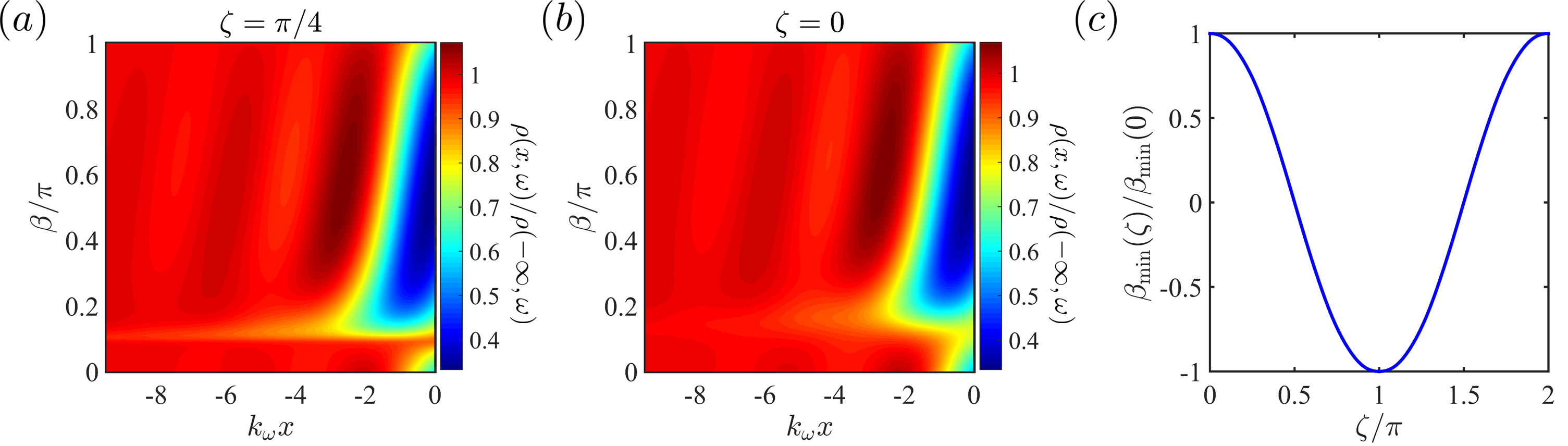}
\caption{LDOS oscillations of the TI-vacuum region shown in Fig.~\ref{schem}, where the TI-SE region is assumed to have a nonhelical interface Hamiltonian exhibiting an elliptical Fermi surface. Figures (a) and (b) showcase the results for the angles $\zeta = \pi/4$ and $\zeta = 0$ respectively. In (c), $\beta_{\rm min}(\zeta)$ is defined as the value of $\beta$ in which the LDOS oscillations disappear for all $x$. When $\zeta = \pi/2$, (which in Appendix~\ref{appendixtransmission} we show is identical to the equal circle case of Section~\ref{sec3}) this is simply $\beta_{\rm min}(\pi/2) = 0$. When $\zeta = 0$, it reaches a maximum value of $\beta_{\rm min}(0) = \pi/8$.}
\label{figzetatable}
\end{figure*}

Using the boundary value matrix of Eq.~\eqref{matrixm} we numerically solve the quasiparticle scattering of the interface and calculate the LDOS for several Fermi surface angles $\zeta$. In Fig.~\ref{figzetatable}(a) and Fig.~\ref{figzetatable}(b) we plot the LDOS as a function of both edge potential strength $\beta$ and position for the angles $\zeta = \pi/4$ and $\zeta = 0$, respectively. We note that in the case of $\zeta = \pi/2$ the scattering of the $x = 0$ junction is identical to that of the equal Fermi surface case in Section~\ref{sec3}, and we obtain the same results as shown in Fig.~\ref{fig2}(a). We demonstrate this analytically in Appendix~\ref{appendixtransmission}, where we calculate the transmission of the junction for $\zeta = \pi/2$ and show its equivalence to the two equal circle system. As $\zeta$ deviates from $\pi/2$, however, the resulting LDOS spectra is modified. While the system is still periodic for edge perturbations $\beta\in[0,\pi)$, the relative pattern of the LDOS oscillations with respect to $\beta$ begins to shift upward, as can be seen in the upwards shift of the blue regions in Figs.~\ref{figzetatable}(a) and (b). In addition to this shift, the character of the LDOS oscillations themselves also change. When $\zeta = 0$ (or equivalently, the ``equal circle" case), the LDOS does not oscillate and is \textit{constant} as $\beta = \beta_{\rm min} = 0$, for all $x$. Here we define $\beta_{\rm min}$ as the value of the edge potential for which the transmission is maximized and consequently the LDOS has no oscillations due to the extremely weak scattering from the edge potential. When $\zeta$ decreases, $\beta_{\rm min}$ beings to shift upward and reaches a maximum of $\beta_{\rm min}/\pi = 1/8$ when $\zeta = 0$. However, at these values of $\beta = \beta_{\rm min}$, the LDOS is \textit{not} constant. For both cases, there is a minimum of the LDOS at the $x = 0$ interface, and the LDOS monotonically rises to $\rho(-\infty,\omega)$ as $x \to -\infty$. This behavior is different to what we have observed in Section.~\ref{sec3}, where the LDOS never monotonically rises as $x\to-\infty$. In Appendix~\ref{appendixtransmission} we also analytically derive the transmission of the junction when $\zeta = 0$ and show that it is maximized as $\beta=\beta_{\rm min} = \pi/8$. This is in strong contrast to the ``equal circle" case, where the scattering of the junction was minimized as $\beta = 0$. For a general angle $\zeta$, the scattering at the $x = 0$ edge is minimized and the LDOS oscillations disappear whenever $\beta = \beta_{\rm min}(\zeta)$. In Fig.~\ref{figzetatable}(c) we plot how $\beta_{\min}(\zeta)$ changes as a function of $\zeta$. From this, we see that the direction of the overall shift of the LDOS oscillation pattern depends on the value of $\zeta$. However, for all angles of the elliptical Fermi surface, we find that the decay envelope of the LDOS oscillations is robust against the details of the nonhelical spin texture and strength of the $x = 0$ edge potential, always taking the value $x ^{-3/2}$.

\section{Discussion and conclusions}
\label{sec6}

In this work we analyzed the effects of step-edge disorder, dissimilar Fermi surfaces, and nonhelical spin textures on the LDOS oscillations of quasiparticle interference patterns in lateral heterostructures. It has earlier been shown that the LDOS oscillations emerging from purely helical systems decay away from the atomic step-edge as $x^{-3/2}$, as opposed to 2DEG systems that decay as $x^{-1/2}$, where $x$ is the distance from the step defect. Strikingly, we find that the decay envelope of the LDOS oscillations is robust to the interfacial disorder that may arise in the TI-SE junction. For all effective models studied for the TI-SE interface that are linear in momentum, we find that this decay power remains $x^{-3/2}$ regardless of the type of TR preserving disorder at the edge.

We find that the qualitative nature of the oscillations result from the linearity of the bands and the TR invariance of the wave functions, while the wave length of the oscillations is defined by the size of the Fermi surface in the TI-vacuum region. Quantitative differences in the amplitude of the oscillations depend on the helicity variations between the two regions and the strength of localized edge potentials at the $x = 0$ boundary. We derived the boundary conditions for wave functions on either side of the $x = 0$ step-edge interface, and found that the quasiparticle scattering of the junction is controlled by two distinct effects: First, the spin texture of the TI-SE interface state, as parametrized by the $c_{ij}$ coefficients of Eq.~\eqref{Ham}. Second, the strength of an edge potential localized at the $x = 0$ step-edge, represented by the parameter $\beta$ in the boundary value matrix of Eq.~\eqref{matrixm}. We then solved for the reflection and transmission of electron scattering at the $x = 0$ boundary and derived the LDOS.

The characteristic energy scale of hexagonal warping for Bi$_{2}$Se$_{3}$ (Bi$_{2}$Te$_{3}$) is $590$ meV~\cite{Kuroda2010} ($260$ meV~\cite{An2012,Fu2009}), while its band gap is $\sim350$ meV~\cite{Kuroda2010} ($\sim250$ meV~\cite{Alpichshev2010}). This implies that, for Bi$_{2}$Se$_{3}$, the transport properties that arise uniquely from the surface states are accurately captured by a linear Hamiltonian, as the energy values at which hexagonal warping affect transport signatures are above its semiconducting gap. As such, the surface state mediated LDOS oscillations are well described by our linear effective model. However, for Bi$_{2}$Te$_{3}$, the in-gap surface states display a relatively stronger hexagonal warping for energies close to, but still below, the semiconducting edge. Therefore, for this material, our model is valid for energies that are sufficiently close to the $\Gamma$-point. For Bi$_{2}$Te$_{3}$ with simple step-edge disorder in particular, these higher energy corrections have earlier been demonstrated to result in larger LDOS oscillation decay envelopes such as $x^{-1}$~\cite{An2012}. The model presented in this paper is thus best suited for Bi$_{2}$Se$_{3}$-SE lateral heterostructures.  

Ref.~\cite{Brems2018} has recently predicted that strain control can be used to manipulate the spin degree of freedom via the spin-orbit coupling in TIs, resulting in spin textures and energy dispersions like those shown in Figs.~\ref{figelipangle}(a) and \ref{figelipangle}(b). Controlling strain in the TI-SE region of the experimental apparatus suggested in Fig.~\ref{schem} will therefore consequently affect the observed LDOS oscillations in the TI-vacuum region. We predict that the observation of decay envelopes different from $x^{-3/2}$ will indicate that the TI-SE interface is not faithfully described by the presence of interface disorder alone.

Our results demonstrate that a careful analysis of the boundary effects on TI interface states is vital in understanding the behaviors of scattering experiments in TIs. In this article we have shown that the boundary value matching of TI interface states is more complex than has been commonly assumed, and that the decay envelope of LDOS oscillations about atomic step-edges is insensitive to the presence of symmetry breaking interfacial disorder in TI-SE junctions. Our work thus constitutes a crucial step towards the characterization of robust signatures associated with TI surface state transport, and in understanding the consequences of symmetry breaking and nonhelical spin textures in TI-based devices.

\begin{acknowledgments}
We thank D. E. Sheehy and I. Vekhter for useful discussions. This work was supported by the U.S. National Science Foundation (NSF) via Grant No. DMR-2213429 (M.M.A.), and through the NSF Partnership for Research and Education in Materials via Grant DMR-1828019 (D.A., D.N.S.). D.A. also acknowledges the support of an Emergence grant from Sorbonne Universit{\'e} (TQCNAA project).
\end{acknowledgments}

\onecolumngrid
\appendix

\section{Self-adjoint boundary conditions}
\label{appendixparticlecurrent}
To arrive at the self-adjoint boundary conditions in a junction containing two Dirac-like systems, we consider
\begin{equation}\label{HL}
H_{\rm L}({\bm p})=-\mu_{\rm L}\sigma_{0}+\sum_{i=x,y,z\atop j=x,y}{a_{ij}\sigma_{i}p_{j}}\; {\rm and} \;H_{\rm R}({\bm p})= - \mu_{\rm R}\sigma_{0}+\sum_{i=x,y,z\atop j=x,y}{c_{ij}\sigma_{i}p_{j}}\;.
\end{equation}
$H_{\rm L}$ ($H_{\rm R}$) describes the left (right) side of the junction, and both Hamiltonians are TR symmetric so long as the $a_{ij}$ and $c_{ij}$ coefficients are all real. Note that at this level $H_{\rm L}$ does not necessarily describe a helical TI-vacuum surface state, but is instead a general effective linear Hamiltonian that is TR invariant. Hermiticity requires that the inner product of the total Hamiltonian $H=H_{\rm L}+H_{\rm R}$ with respect to any wave function in our Hilbert space must satisfy $\langle\psi_{1}|H\psi_{2}\rangle=\langle H\psi_{1}|\psi_{2}\rangle$. Since translational invariance is broken only along the $x$-axis, we only need to consider the terms in $H$ that depend on the momentum operator $p_{x}=-i\partial_{x}$. (Here and throughout these Appendices we set $\hbar = 1$). Labeling this part of the Hamiltonain as $H_{x}$, we have that
\begin{equation}\label{hermt2}
\langle\psi_{1}|H_{x}\psi_{2}\rangle=\int_{-\infty}^{0}\psi^{\dag}_{1,{\rm L}}(x)\left(\sum_{i=x,y,z}{a_{ix}\sigma_{i}(-i\partial_{x})}\right)\psi_{2,{\rm L}}(x)dx+\int_{0}^{\infty}\psi^{\dag}_{1,{\rm R}}(x)\left(\sum_{i=x,y,z}{c_{ix}\sigma_{i}(-i\partial_{x})}\right)\psi_{2,{\rm R}}(x)dx\;.
\end{equation}
Integrating by parts and assuming the wave functions are well-behaved at infinity, this implies that $\langle\psi_{1}|H\psi_{2}\rangle=\langle H\psi_{1}|\psi_{2}\rangle$ is satisfied if
\begin{equation}\label{cond1}
\psi^{\dag}_{1,{\rm L}}(0)\left(\sum_{i=x,y,z}{a_{ix}\sigma_{j}}\right)\psi_{2,{\rm L}}(0)=\psi^{\dag}_{1,{\rm R}}(0)\left(\sum_{i=x,y,z}{c_{ix}\sigma_{j}}\right)\psi_{2,{\rm R}}(0)\;.
\end{equation}
Since this condition must hold for any pair of wave functions in the Hilbert space we can solve it by demanding
\begin{equation}\label{inteface1}
\psi_{\rm L}(0)=\mathcal{M}\psi_{\rm R}(0)
\end{equation}
at the interface. Here $\mathcal{M}$ is a $2\times 2$ matrix with arbitrary elements. If we substitute Eq.~\eqref{inteface1} in Eq.~\eqref{cond1}, we obtain
\begin{equation}\label{cond3}
\sum_{i=x,y,z}{c_{ix}\sigma_{i}}=\mathcal{M}^{\dag}\left(\sum_{i=x,y,z}{a_{ix}\sigma_{i}}\right)\mathcal{M}\;.
\end{equation}
For linearly dispersing systems, the self-adjoint boundary condition as expressed by Eq.~\eqref{cond3} can also be obtained through the consideration of current conservation across the junction.  The $x$-component of the current in the left and right Dirac-like systems are given by
\begin{equation}\label{currents}
j_{{\rm L},x}=\psi^{\dag}_{\rm L}\left(\sum_{i=x,y,z}{a_{ix}\sigma_{i}}\right)\psi_{\rm L}\; {\rm and }\; j_{{\rm R},x}=\psi^{\dag}_{\rm R}\left(\sum_{i=x,y,z}{c_{ix}\sigma_{i}}\right)\psi_{\rm R}\;.
\end{equation}
Making use of the boundary condition in Eq.~\eqref{inteface1}, we recover the expression in Eq.~\eqref{cond3}. This shows that for linearly dispersing systems both Hermiticity and current conservation lead to equivalent self-adjoint boundary conditions.
\subsection{Discrete symmetries}
\label{appendixdiscretesymmetries}
While the boundary matrix condition obtained in Eq.~\eqref{cond3} is generic, its form can be constrained by physical considerations. Discrete symmetries such as time reversal (TR) symmetry $\mathcal{T} = i\sigma_{y}K$ ($K$ being the complex conjugation operator), particle hole (PH) symmetry $\mathcal{P} = \sigma_{x}K$, and chiral symmetry $\mathcal{C} = \mathcal{T}\mathcal{P}$ impose restrictions on the matrix elements introduced in Eq.~\eqref{inteface1}. To demonstrate this, let us suppose that the system possesses an arbitrary discrete symmetry defined by $\mathcal{O}$. For any wave function $\psi_{1}$ of our Hilbert space $\mathbb{H}$, there exists a wave function $\psi_{2} = \mathcal{O}\psi_{1}$ such that $\psi_{2}\in\mathbb{H}$. Because Eq.~\eqref{inteface1} must hold for all wave functions in the Hilbert space, it follows that
\begin{eqnarray}\label{1and2}
  \psi_{1,{\rm L}}(0) =& \mathcal{M}\psi_{1,{\rm R}}(0)\; {\rm and}\; \psi_{2,{\rm L}}(0) = \mathcal{M}\psi_{2,{\rm R}}(0)\;.
\end{eqnarray}
If we apply the operator $\mathcal{O}$ to the first equation in Eq.~\eqref{1and2}, we obtain
\begin{eqnarray}
  \mathcal{O}\psi_{1,{\rm L}}(0) &=& \mathcal{O}\mathcal{M}\psi_{1,{\rm R}}(0)\;.
\end{eqnarray}
Comparing this to the second equation in Eq.~\eqref{1and2}, we observe that $[\mathcal{M},\mathcal{O}]=0$. We note that while $\mathcal{T}$, $\mathcal{P}$, and $\mathcal{C}$ all commute with the boundary value matrix $\mathcal{M}$, these discrete symmetries need not all commute with the Hamiltonians $H_{\rm L}$ and $H_{\rm R}$. For TR symmetric systems, we obtain $[H_{\rm L,R},\mathcal{T}] = 0$. However, for PH and Chiral symmetric systems, we obtain $\{H_{\rm L,R},\mathcal{P}\} = \{H_{\rm L,R},\mathcal{C}\} = 0$.
\subsection{TR invariant self-adjoint boundary conditions}
\label{appendixgammas}
Consider the following TR invariant Hamiltonians
\begin{equation}\label{HL2}
H_{\rm L}({\bm p})=v_{\rm F}({\bm \sigma}\times {\bm p })_{z} - \mu_{\rm L}\; {\rm and} \; H_{\rm R}({\bm p})=\sum_{i=x,y,z\atop j=x,y}{c_{ij}\sigma_{i}p_{j}} - \mu_{\rm R}\;.
\end{equation}
In this case $\mathcal{M}$ satisfies
\begin{equation}\label{cond332g}
\sum_{i=x,y,z}c_{ix}\sigma_{i}=-v_{\rm F}\mathcal{M}^{\dag}\sigma_{y}\mathcal{M}\;.
\end{equation}
Moreover, if we assume that the system is TR invariant it follows that $[\mathcal{M},\mathcal{T}]=0$. This condition constrains the matrix elements of $\mathcal{M}$ such that
\begin{equation}\label{mmtrs2g}
\mathcal{M}=\gamma_{0}\sigma_{0}+i\sum_{i=x,y,z}{\gamma_{i}\sigma_{i}}\;,
\end{equation}
where $\gamma_{0},\gamma_{x},\gamma_{y},\gamma_{z}\in \mathbb{R}$ (here $\mathbb{R}$ is the set of real numbers). Substituting Eq.~\eqref{mmtrs2g} in Eq.~\eqref{cond332g} and solving for the coefficients we find
\begin{equation}\label{solustions2g}
(\gamma_{0},\gamma_{x},\gamma_{y},\gamma_{z})=\begin{cases}
\left(\gamma_{0},\frac{|{\bm c}_{x}| - c_{yx}}{c^{2}_{xx}+c^{2}_{zx}}\left[c_{zx}\gamma_{0}\pm c_{xx}\gamma_{y}\right], \pm\sqrt{-\left(\frac{ |{\bm c}_{x}| + c_{yx} }{2v_{\rm F}}+ \gamma^{2}_{0}\right)},\frac{|{\bm c}_{x}|-c_{yx}}{c^{2}_{xx}+c^{2}_{zx}} \left[c_{xx}\gamma_{0}\mp c_{zx}\gamma_{y}\right] \right),\\
\left(\gamma_{0},\frac{|{\bm c}_{x}| + c_{yx}}{c^{2}_{xx}+c^{2}_{zx}}\left[\pm c_{zx}\gamma_{0}+c_{xx}\gamma_{y}\right], \pm\sqrt{\frac{|{\bm c}_{x}| - c_{yx} }{2v_{\rm F}} - \gamma^{2}_{0}},\frac{|{\bm c}_{x}| + c_{yx}}{c^{2}_{xx}+c^{2}_{zx}}\left[c_{xx}\gamma_{0}\mp c_{zx}\gamma_{y}\right] \right)\;.
\end{cases}
\end{equation}
Here we have defined the vector
\begin{equation}
{\bm c_{i}}= c_{xi}\widehat{\bm{x}} + c_{yi}\widehat{\bm{y}} + c_{zi}\widehat{\bm{z}}\;,
\end{equation}
and thus $|{\bm c_{i}}|=\sqrt{c^{2}_{xi}+c^{2}_{yi}+c^{2}_{zi}}$. Note that this definition of $\bm{c}_{i}$ is distinct from the definitions of $\bm{c}$, $c_{i}$, $\bm{c}(\bm{k})$, and $c_{i}(\bm{k})$ given earlier in the manuscript. We discard the first set of solutions in Eq.~\eqref{solustions2g} since in order to have $\gamma_{x,y,z}\in \mathbb{R}$, the remaining coefficient $\gamma_{0}$ must be purely imaginary and thus breaks TR. The second set of solutions in Eq.~\eqref{solustions2g} satisfies TR (i.e. $\gamma_{0},\gamma_{x},\gamma_{y},\gamma_{z}\in \mathbb{R}$) if
\begin{equation}\label{condd2g}
-\sqrt{\frac{|{\bm c_{x}}|-c_{yx}}{2v_{\rm F}}}\leq\gamma_{0}\leq\sqrt{\frac{|{\bm c_{x}}|-c_{yx}}{2v_{\rm F}}}\;, \; {\rm equivalently},\; \gamma_{0}=\sqrt{\frac{|{\bm c_{x}}|-c_{yx}}{2v_{\rm F}}}\cos(\beta)\;,
\end{equation}
where $\beta$ is a free real parameter. Hence the boundary matrix that grants the TR constraint is given by
\begin{equation}\label{newmat2h2}
\mathcal{M}=\sqrt{\frac{|{\bm c_{x}}|-c_{yx}}{2v_{\rm F}}}e^{i\sigma_{y}\beta}+\frac{i}{\sqrt{2v_{\rm F}(|{\bm c_{x}}|-c_{yx})}}\left(c_{xx}\sigma_{z}-c_{zx}\sigma_{x}\right)e^{-i\sigma_{y}\beta}\;.
\end{equation}
\section{Scattering in lateral junctions}
\label{Appendixscatteringdetails}
If we consider the junction described by Eq.~\eqref{HL2} and if we assume an incoming plane wave from the left, then the spinor wave function for $x<0$ can be written as an incoming and reflected electron
\begin{equation}\label{leftside}
\psi_{\rm L}({\bm r})= \dfrac{e^{i(k_{x}x+k_{y}y)}}{\sqrt{2}} \left(
               \begin{array}{c}
                 1 \\
                -i  e^{i\theta} \\
               \end{array}
             \right)+ r \dfrac{e^{i(-k_{x}x+k_{y}y)}}{\sqrt{2}} \left(
               \begin{array}{c}
                 1 \\
                 i e^{-i\theta} \\
               \end{array}
             \right)\;,
\end{equation}
where $\bm{k} = (k_{x},k_{y}) = |\bm{k}|[\cos(\theta),\sin(\theta)]$ is the momentum of the incoming plane wave, $\theta=\tan^{-1}(k_{y}/k_{x})$,
 and $r$ is the reflection coefficient. Considering that translational invariance along the $y$-direction requires the conservation the $y$-component of the momentum we can express the electrons transmitted to the right as
\begin{equation}\label{psirg}
  \psi_{\rm R}({\bm r})=te^{i(k'_{x}x+k_{y}y)}\left(
                      \begin{array}{c}
                        \dfrac{ \sqrt{ \epsilon({\bm k'})+h_{z}({\bm k'})} }{ \sqrt{2\epsilon(\bm{k}')} } \\
                      \dfrac{  h_{x}({\bm k'})+ih_{y}({\bm k'}) }{ \sqrt{2\epsilon(\bm{k}')} \sqrt{ \epsilon({\bm k'})+h_{z}({\bm k'})} } \\
                      \end{array}
                    \right)\;,
\end{equation}
where $t$ is the transmission coefficient. Here we have defined
\begin{equation}\label{er}
  \epsilon({\bm k'})=\sqrt{\sum_{i=x,y,z}{h^{2}_{i}({\bm k'})}}\; {\rm and }\; h_{i}({\bm k'})= c_{ix}k_{x}' + c_{iy}k_{y}\;,
\end{equation}
with ${\bm k'}=(k'_{x},k_{y})$. Additionally, the conservation of energy $ v_{\rm F}\sqrt{k^{2}_{x}+k^{2}_{y}} - \mu_{\rm L} = \epsilon(\bm{k}') - \mu_{\rm R}$ along with the requirement that the momentum must lie on the Fermi surface as $v_{\rm F}\sqrt{k_{x}^{2} + k_{y}^{2}} - \mu_{\rm L} = 0$ allows us to find the $k'_{x}$ component of the transmitted momentum, i.e.
\begin{equation}\label{kxpg}
k'_{x}=\frac{-{\bm c_{x}}\cdot{\bm c_{y}}k_{y}+\sqrt{({\bm c_{x}}\cdot{\bm c_{y}}k_{y})^2 + |{\bm c_{x}}|^{2}(\mu_{\rm R}^{2} - |{\bm c_{y}}|^{2}k^{2}_{y})}}{|{\bm c_{x}}|^{2}}\;.
\end{equation}
Now that we have determined the outgoing momenta in Eq.~\eqref{psirg} we can find the coefficients $r$ and $t$ [Eqs.~\eqref{leftside} and \eqref{psirg}] through the boundary condition at $x=0$, $\psi_{\rm L}(0)=\mathcal{M}\psi_{\rm R}(0)$, and the matrix $\mathcal{M}$ in Eq.~\eqref{newmat2h2}. The boundary value problem can be expressed by the matrix equation
\begin{equation}\label{bcondmatrix}
\left(
  \begin{array}{cc}
    -1 & \nu[h_{x}({\bm k'})+ih_{y}({\bm k'})]+\eta[\epsilon +h_{z}({\bm k'})] \\
    -i e^{-i\theta} &  \eta^{*}[h_{x}({\bm k'})+ih_{y}({\bm k'})]-\nu^{*}[\epsilon +h_{z}({\bm k'})]  \\
  \end{array}
\right)\left(
                                                                                                                \begin{array}{c}
                                                                                                                  r \\
                                                                                                                  t [\epsilon^{2} + \epsilon h_{z}]^{-1/2} \\
                                                                                                                \end{array}
                                                                                                              \right)=\left(
                                                                                                                \begin{array}{c}
                                                                                                                  1 \\
                                                                                                                  -i e^{i\theta} \\
                                                                                                                \end{array}
                                                                                                              \right)\;.
\end{equation}
In Eq.~\eqref{bcondmatrix} we have defined
\begin{equation}\label{eta}
  \eta = \kappa\cos(\beta)+i\left(\frac{c_{xx}\cos(\beta)-c_{zx}\sin(\beta)}{2v_{\rm F}\kappa}\right)\;,
  \nu = \kappa\sin(\beta)-i\left(\frac{c_{zx}\cos(\beta)+c_{xx}\sin(\beta)}{2v_{\rm F}\kappa}\right)\;, {\rm and}\;
  \kappa = \sqrt{\frac{|{\bm c_{x}}|-c_{yx}}{2v_{\rm F}}}\;.
\end{equation}
Solving for $r$ we find
\begin{equation}\label{rgencase}
r(\theta)=\frac{  e^{i
   \theta } \left[ \eta (\epsilon+h_z)+\nu
   (h_x+i
   h_y)\right]+ i\left[\nu^{*}
   (\epsilon+h_z) - \eta^{*}
   (h_x + i h_y)
   \right] }{ e^{-i\theta} \left[ \eta
   (\epsilon+h_z)+\nu
   (h_x+i h_y)\right] - i
   \left[ \nu^{*}
   (\epsilon+h_z) - \eta^{*}
   (h_x+i h_y) \right]}\;.
\end{equation}
In order to define the reflection and transmission probability amplitudes we use the fact that the current in the left side of the junction must be equal to the current in the right side of the junction
\begin{equation}\label{current}
j_{x}[{\rm incoming}]+j_{x}[{\rm reflected}]=j_{x}[{\rm transmitted }]\;, {\rm i.e. }\;, 1=\frac{j_{x}[{\rm transmitted }]}{j_{x}[{\rm incoming}]}-\frac{j_{x}[{\rm reflected}]}{j_{x}[{\rm incoming}]}=T(\theta)+R(\theta)\;,
\end{equation}
where $T(\theta)$ [$R(\theta)$] are the transmission (reflection) probability amplitudes. Making use of the incoming, reflected, transmitted states, Eqs.~\eqref{leftside} and \eqref{psirg}, and the definition of the currents in Eq.~\eqref{currents}, we obtain
\begin{equation}\label{rtone}
 R(\theta)= -\frac{j_{x}[{\rm reflected}]}{j_{x}[{\rm incoming}]}=|r(\theta)|^{2}\;{\rm and } \;T(\theta)=\frac{j_{x}[{\rm transmitted }]}{j_{x}[{\rm incoming}]}=\frac{\psi_{\rm R}^{\dag}\left(\sum_{i=x,y,z}{c_{ix}\sigma_{i}}\right)\psi_{\rm R}}{v_{\rm F}\cos\theta}=1-R(\theta)
\end{equation}
since $j_{x}[{\rm incoming}]=v_{\rm F}\cos(\theta)$ and $j_{x}[{\rm reflected}]=-v_{\rm F}|r(\theta)|^{2}\cos(\theta)$~\cite{Allain2011}.
\subsection{Special cases}
Here we analyze the transmission amplitudes for two distinct cases:
\begin{equation}\label{HamRcases}
 H_{{\rm R}1} =\lambda_{1} v_{\rm F}({\bm \sigma}\times {\bm p})_{z} - \mu\;{\rm and } \;H_{{\rm R}2} = v_{\rm F}\left[({\bm \sigma}\times {\bm p})_{z}+\lambda_{2}({\bm \sigma}\cdot \bar{\bm{e}})({\bm p}\cdot{\bm e})-\lambda_{3}\sigma_{z}{\bm p}\cdot{\bm e}\right] - \mu.\;
\end{equation}
In the following we set $\mu_{\rm L} = \mu_{\rm R} = \mu$. The first case corresponds to a system with a different Fermi velocity. The second case in nonhelical and corresponds to a system with no inversion symmetry, resulting an elliptical Fermi surface and spins that point out of the $x-y$ plane. In this case, the unit vectors $\bm{e} = [\sin(\zeta),\cos(\zeta),0]$ and $\bar{\bm{e}} = \widehat{\bm{z}}\times  \bm{e}  = [-\cos(\zeta),\sin(\zeta),0]$ are defined to be within the $x-y$ plane, and point along the semi-major and semi-minor axes of the elliptical Fermi surface respectively, with the angle $\zeta$ being defined from the $y$-axis. Details of their description can be found in Ref.~\cite{Asmar2017}.

\subsubsection{Two different velocities}
\label{appendixreflection}
Consider a junction described by $H_{\rm L}=v_{\rm F}({\bm \sigma}\times {\bm p })_{z} - \mu$ and $ H_{{\rm R}1}=\lambda_{1} v_{\rm F}({\bm \sigma}\times {\bm p})_{z} - \mu$, where $0<\lambda_{1}\le 1$. The incoming and reflected states for $x<0$ are given in Eq.~\eqref{leftside}. The outgoing momentum can be expressed as ${\bm k'}=|\bm{k}'|[\cos(\theta'),\sin(\theta')]$, where $\theta'=\tan^{-1}(k_{y}/k'_{x})$
 is the transmitted angle. Since the momentum along $y$ is conserved the relation between the incoming and transmitted angles is given by $\sin(\theta')/\lambda_{1}=\sin(\theta)$ (electronic Snell's law). For $H_{{\rm R}1}$ we have $c_{yx}=-c_{xy}=-\lambda_{1}v_{\rm F}$ and $c_{xx}=c_{yy}=c_{zx}=c_{zy}=0$, giving us $h_{x}(\bm k')=\epsilon(\bm k')\sin(\theta')$, $h_{y}(\bm k')=-\epsilon(\bm k')\cos(\theta')$, and  $\epsilon({\bm k}')=\lambda_{1}v_{\rm F}\sqrt{k^{2}_{y}+k^
{'2}_{x}}$ [Eq.~\eqref{er}]. The transmitted states in Eq.~\eqref{psirg} are then given by
\begin{equation}\label{lefts}
\psi_{\rm R}({\bm r})=t \dfrac{e^{i(k'_{x}x+k_{y}y)}}{\sqrt{2}} \left(
               \begin{array}{c}
                 i \\
                 e^{i\theta'} \\
               \end{array}
             \right)\;.
\end{equation}
 Moreover, the $x$-component of the transmitted momentum in Eq.~\eqref{kxpg} reduces to $k'_{x}=[\mu/(v_{\rm F}\lambda_{1})]\sqrt{1-[\lambda_{1}\sin(\theta)]^2}$. Hence, with $\theta'$ and $k'_{x}$ we can find $r(\theta)$ in Eq.~\eqref{rgencase} to be
\begin{equation}\label{rviv2}
r(\theta)= \frac{e^{i\theta}\left[i\sin(\frac{\theta-\theta'}{2})\cos(\beta)+\sin(\frac{\theta+\theta'}{2})\sin(\beta)\right]}{\cos(\frac{\theta+\theta'}{2})\cos(\beta)-i\cos(\frac{\theta-\theta'}{2})\sin(\beta)}\;.
\end{equation}
From $r(\theta)$ the transmission probability amplitude $T(\theta)=1-|r(\theta)|^{2}$ is given by
\begin{equation}\label{Tv1v2}
T(\theta)=\frac{\cos(\theta')\cos(\theta)}{\cos^{2}\left(\frac{\theta+\theta'}{2}\right)\cos^{2}(\beta)+\cos^{2}\left(\frac{\theta-\theta'}{2}\right)\sin^{2}(\beta)}=\frac{2\cos(\theta)\sqrt{1-\lambda^{2}_{1}\sin^{2}(\theta)}}{1-\cos(2\beta)\sin^{2}(\theta)+\cos(\theta)\sqrt{1-\lambda^{2}_{1}\sin^{2}(\theta)}}\;.
\end{equation}
\subsubsection{Nonhelical spin textures}
\label{appendixtransmission}
In this section we shall analyze a generalized version of Eq.~\eqref{cperp}. For a system described by $H_{\rm L}=v_{\rm F}({\bm \sigma}\times {\bm p })_{z} - \mu$ and $ H_{{\rm R}2}$, where $0<\lambda_{2,3}\le 1$, the incoming and reflected states for $x<0$ are given in Eq.~\eqref{leftside}. For $H_{{\rm R}2}$ we have ${\bm c}_{x}=(c_{xx},c_{yx},c_{zx})=v_{\rm F}[-\lambda_{2}\cos(\zeta)\sin(\zeta),-1+\lambda_{2}\sin^{2}(\zeta),-\lambda_{3}\sin(\zeta)]$ and ${\bm c}_{y}=(c_{xy},c_{yy},c_{zy})=v_{\rm F}[1-\lambda_{2}\cos^{2}(\zeta),\lambda_{2}\cos(\zeta)\sin(\zeta),-\lambda_{3}\cos(\zeta)]$. Since $k_{y}$ is conserved we have $h_{x}({\bm k'}) = v_{\rm F}\left(k_{y}-\lambda_{2}\cos(\zeta)[k_{y}\cos(\zeta)+k'_{x}\sin(\zeta)]\right)$, $h_{y}({\bm k'}) = v_{\rm F}\left(-k'_{x}+\lambda_{2}\sin(\zeta)[k_{y}\cos(\zeta)+k'_{x}\sin(\zeta)]\right)$, and $h_{z}({\bm k'})=-v_{\rm F}\lambda_{3}[k_{y}\cos(\zeta)+k'_{x}\sin(\zeta)]$. Additionally, $k'_{x}$ can be found from energy conservation such that
\begin{equation}\label{kxpb2}
k'_{x}=\dfrac{k \left(\sqrt{4 \cos ^2(\theta )-2 \left[(\lambda_{2}-2)
   \lambda_{2}+\lambda_{3}^2\right] \left[\cos (2 \zeta )-\cos (2 \theta
   )\right]}-\sin (2 \zeta ) \sin (\theta ) \left[(\lambda_{2}-2) \lambda_{2}+\lambda_{3}^2\right]\right)}{-\cos (2 \zeta ) \left[(\lambda_{2}-2) \lambda_{2}+\lambda_{3}^2\right]+(\lambda_{2}-2)
   \lambda_{2}+\lambda_{3}^2+2}\;.
\end{equation}
Here $k = |\bm{k}| = \sqrt{k_{x}^{2} + k_{y}^{2}}$. Now by substituting in Eq.~\eqref{rgencase} we can write $r(\theta)$ for arbitrary values of $\zeta$.

In order to gain physical insight into the dependence of the transmission on $\zeta$ we explore two limiting cases. First we set $\zeta=\pi/2$, i.e. the axis of conserved reflections is parallel to the lateral-edge, and the major axis of the elliptical Fermi surface points along the $x$-direction. In this case $ h_{x}({\bm k'})= v_{\rm F}k_{y}$, $h_{y}({\bm k'}) = v_{\rm F} \left(-1+\lambda_{2}\right)k'_{x}$, $h_{z}({\bm k'}) = -v_{\rm F} \lambda_{3}k'_{x}$, and $k'_{x}$ reduces to  $k'_{x}=k_{x}/\sqrt{(\lambda_{2}-1)^2+\lambda^{2}_{3}}$. Hence, by substituting in Eq.~\eqref{rgencase} and calculating $T(\theta)=1-|r(\theta)|^{2}$, where here $r(\theta)$ is equal to Eq.~\eqref{equalcircler}, we obtain
\begin{equation}\label{a2trnasitFIRST}
T(\theta)=\frac{\cos^{2}(\theta)}{\cos^{2}\left(\theta\right)\cos^{2}(\beta)+\sin^{2}(\beta)}\;.
\end{equation}
It is important to notice that in this case $T(\theta)$ becomes identically unity as $\beta=0$, i.e., when there is no interface scattering, and displays perfect transparency for all incoming angles. Moreover, we can notice that Eq.~\eqref{a2trnasitFIRST} is identical to Eq.~\eqref{Tv1v2} for $\theta'=\theta$, which is satisfied whenever $\lambda_{1}=1$, i.e., equal Fermi velocities. Additionally, setting $\theta'=\theta$ in Eq.~\eqref{rviv2} allows us to recover Eq.~\eqref{equalcircler}. This is to say, the transmission in this $\zeta = \pi/2$ case is identical to that of a junction of two helical and equal-sized Fermi surfaces.

Second, if $\zeta=0$ the axis of reflection of the system is orthogonal to the lateral-edge, and the major axis of the elliptical Fermi surface points along the $y$-direction. In this case $h_{x}({\bm k'}) = v_{\rm F}\left(1-\lambda_{2}\right)k_{y}$, $ h_{y}({\bm k'}) = -v_{\rm F}k'_{x}$, $h_{z}({\bm k'}) = -v_{\rm F}\lambda_{3}k_{y}$, and $k'_{x}=k\sqrt{1-\left[(\lambda_{2}-1)^2+\lambda^{2}_{3}\right] \sin^{2}(\theta)}$, and we find
\begin{equation}\label{a2trnasit}
T(\theta)=\frac{2 \cos (\theta ) \sqrt{1-\sin ^2(\theta ) \left[(\lambda_{2}-1)^2+\lambda_{3}^2\right]}}{\sin ^2(\theta ) \left[(\lambda_{2}-1) \cos (2 \beta )-\lambda_{3} \sin (2 \beta )\right] +\cos (\theta )
   \sqrt{1-\sin ^2(\theta ) \left[(\lambda_{2}-1)^2+\lambda_{3}^2\right]}+1}\;.
\end{equation}
For brevity, in this case we only provide $T(\theta)$. Unlike the previous cases, we notice that maximum transmission does not occur at $\beta=0$. Here the value of $\beta$ that maximizes the transmission ($\beta_{max}$) satisfies the condition $\lambda_{3}\cos(2\beta_{max})+(\lambda_{2}-1)\sin(2\beta_{max})=0$. Hence in this case $\beta_{max}=\frac{1}{2}\tan^{-1}\left(\frac{\lambda_{3}}{1-\lambda_{2}}\right)$. Thus, when $\lambda_{2} = 1/3$ and $\lambda_{3} = 2/3$ as is the case for Eq.~\eqref{cperp}, we have that $\beta_{max} = \pi/8$. This results in the LDOS given in Fig.~\ref{figzetatable}(b).

\twocolumngrid

\bibliography{bibfile}

\bibliographystyle{apsrev4-2}

\end{document}